\documentclass[aps,amssymb,prl,twocolumn,showpacs,amsmath]{revtex4-1} \fussy

\usepackage{times}
\usepackage[usenames]{color}
\usepackage{graphicx,amssymb,amsmath} %% ensures you can use EPS pictures
\usepackage{bm}

\begin{document}
\title{Ubiquity of optical activity in planar metamaterial scatterers}
\author{Ivana Sersic}\email{i.sersic@amolf.nl}\homepage{http://www.amolf.nl/research/resonant-nanophotonics/}
\affiliation{Center for Nanophotonics, FOM Institute for Atomic and Molecular Physics (AMOLF), Science Park 104, 1098 XG Amsterdam, The Netherlands}
\author{Marie Anne van de Haar}
\author{Felipe Bernal Arango}
\author{A. Femius Koenderink}
\affiliation{Center for Nanophotonics, FOM Institute for Atomic and Molecular Physics (AMOLF), Science Park 104, 1098 XG Amsterdam, The Netherlands}
\begin{abstract}
Recently it was discovered that periodic lattices of metamaterial scatterers show optical activity, even if the scatterers or lattice show no 2D or 3D chirality, if the illumination breaks symmetry. In this Letter we demonstrate that such `pseudo-chirality' is intrinsic to any single planar metamaterial scatterer and in fact has a well-defined value at a universal bound. We argue that in any  circuit model,  a nonzero electric and magnetic polarizability derived from a single resonance automatically imply   strong bianisotropy, i.e., magneto-electric cross polarizability at the universal bound set by energy conservation. We confirm our claim  by extracting  polarizability tensors and cross sections for handed excitation   from  transmission measurements on near-infrared split ring arrays, and electrodynamic simulations for diverse  metamaterial scatterers.
\end{abstract}
\date{Posted on arxiv Jan. 13, 2012}
\maketitle

%paragraph{Introduction}
Many historical  debates on how to describe the effective electrodynamic response of  media composed of  subwavelength building blocks  currently acquire new relevance in nano-optics. Initiated by the works of Veselago and Pendry~\cite{Veselago68,Pendry00}, efforts are focused on manipulating effective medium parameters in nanostructured media. On the one hand, the drive for  arbitrary $\epsilon$ and $\mu$  is generated by  the idea that light fields can be arbitrarily reshaped by conformal transformations, provided  we can create arbitrary constitutive tensors~\cite{Pendry06,Leonhardt06,Schurig06}.  On the other hand, a convergence with plasmonics has led to the realization that  subwavelength scatterers mimic and even greatly enhance rich scattering phenomena known from molecular matter. For example, resonantly induced optical magnetism in 2D and 3D chiral metal nano-objects results in   giant  circular birefringence, optical rotatory power, broadband    optical activity, and circular dichroism in frequency ranges from  microwave, mid-IR, near IR to even visible frequencies~\cite{Valev09,Gorodetsky09,Gansel09,Drezet08,Zhang09,Decker10,Li10,Zhou09,Plum09}.
The fact that strong optical activity is easily attained using   chiral subwavelength scatterers is promising for many   applications such as broadband optical components,  as well as providing excellent candidates for achieving negative refraction~\cite{Pendry04}, or repulsive Casimir forces~\cite{SoukoulisCasimir}. Moreover, the promise of enhancing detection of molecular chirality via enhanced chirality in the excitation field, is expected to be of large importance for, e.g.,  discrimination of enantiomers in biology or medicine~\cite{Govorov10,Hendry10,Cohen10,Cohen11}.

\begin{figure}
\includegraphics[width=\columnwidth]{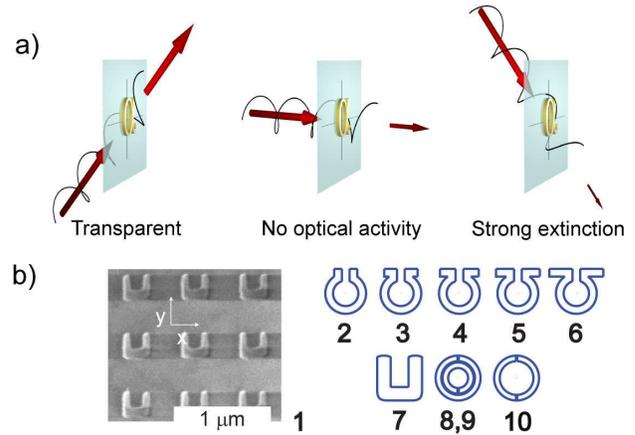}
\caption{a) Any scatterer $\alpha$  with nonzero electric and magnetic polarizability shows oblique incidence optical activity, with transparency for one handedness of incident light at off-angles, and maximum extinction when the incident beam is rotated by 90$^\circ$. At normal incidence, the scatterer shows no optical activity. b) Common planar  scatterers for which we verify optical activity and bianisotropy: (1) scanning electron micrograph of 230$\times$30$\times$30 nm Au SRRs. Structures (2)-(6): $\Omega$ particles of varying arm length. Structure (7) model for SRR in (1). Structure (8,9,10): double split ring and double gap ring~\cite{Kern09}. }
\label{cartoon}
\end{figure}

A question of essential importance is how  to control  the optical activity of a single building block, i.e.,  have independent control over the degree of magnetic response, electric response  and magneto-electric cross coupling or `bianisotropy' whereby incident electric (magnetic) fields cause a magnetic (electric) material polarization in a single building block~\cite{SihvolaBook}.  For instance,  in attempts to reach negative indices, researchers soon found that the  archetypical split ring resonator (SRR) has  a magneto-electric response that is undesirable, yet difficult to remove without also losing the magnetic response~\cite{SoukoulisAPL}.   Completely opposite to the desire to remove this bianisotropy, it has also been realized that all applications exploiting optical activity benefit from strong magneto-electric coupling.  Currently it is unclear if there exists any  universal bound  to which optical activity can be benchmarked, or conversely, if it is at all possible to avoid bianisotropy without also losing the magnetic response~\cite{Sersic11a}.
In this Letter, we discuss precisely such a universal bound for   magneto-electric coupling for single scatterers, disentangled from any lattice properties. We claim that  Onsager's relations constrain optical activity to always be at this maximum bound  for any dipole scatterer based on planar circuit designs, independent of geometrical chirality.  Our claim is supported by measurements on SRRs  at telecom wavelengths and rigorous full wave calculations~\cite{Kern09} in which we retrieve  cross sections  and polarizabilities  for various metamaterial scatterers (see Fig.~\ref{cartoon}(a,b)).

The central quantity in this Letter is the polarizability tensor that quantifies  the magnetic response, electric response and magneto-electric cross coupling (bianisotropy)  intrinsic to a  single metamaterial building block according to~\cite{SihvolaBook,Sersic11a}:
\begin{equation}
\begin{pmatrix}\mathbf{p}\\ \mathbf{m} \\ \end{pmatrix} = \begin{pmatrix}\mathbf{\alpha}_E &  i\mathbf{\alpha}_C\\   -i\mathbf{\alpha}^T_C  & \mathbf{\alpha}_H \\ \end{pmatrix}
 \begin{pmatrix}\mathbf{E}\\ \mathbf{H} \end{pmatrix}\label{eq:generalform}
\end{equation}
For molecules, optical activity is due to weak  cross coupling, i.e., a perturbative $\alpha_C H \approx 10^{-3} \alpha_E E$, while $\alpha_H\approx 0$. In contrast, the paradigm of metamaterials is that a   single scatterer acquires   a magnetic  dipole moment $\mathbf{m}$ at least comparable to the electric moment  $\mathbf{p}$, with  $\alpha_E$, $\alpha_H$, and possibly $\alpha_C$ of the same order, which all derive from a single resonance~\cite{footnote}.
\begin{figure}
\includegraphics[width=\columnwidth]{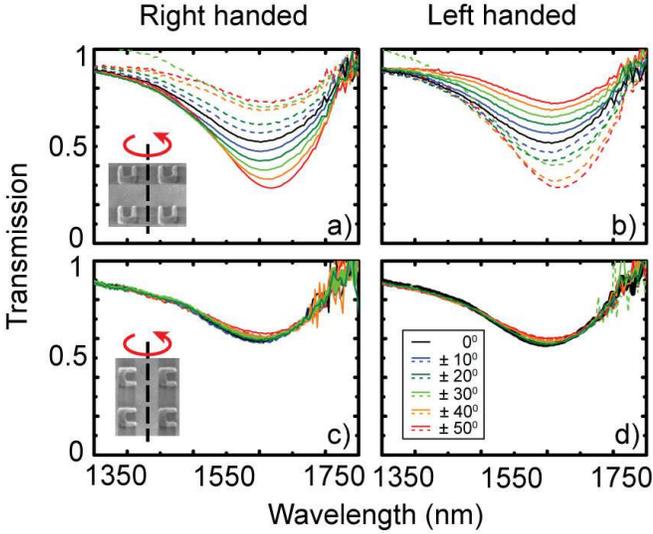}
\caption{Transmission spectra from a periodic square array of 230x230 nm SRR with $d$=530 nm. The spectra were taken as a function of angle of incidence, where dashed curves denote negative angles, and solid curves positive angles with respect to the sample normal. (a,c) and (b,d) are transmission spectra shown for right- and left-handed circularly polarized illumination. Inset in (a)  resp.~(c) shows the  the sample rotation axis for (a,b) resp.~(c,d), taking the incident $k$-vector as pointing through the paper.} \label{measurements}
\end{figure}
In order to quantify the polarizability for the canonical SRR, we performed transmission measurements as well as full-wave calculations.
For the experiments we  fabricated  Au SRRs arranged in square arrays on glass substrates by   electron beam lithography (e-beam), resonant at telecom wavelengths~\cite{Enkrich05,Sersic09}. Fig.~\ref{cartoon}(b) shows a scanning electron micrograph (SEM) of a  SRR array with 530 nm lattice spacing,  which is so dilute that coupling between SRRs is   small~\cite{Sersic09}, yet so dense that no grating diffraction occurs.  Each  SRR measures 230$\times$230$\times$30 nm, with a gap between the arms that is 100 nm wide and 145 nm deep.  We record transmission by illuminating the sample with a narrow band of  frequencies at a time, selected from a supercontinuum laser (Fianium), using an acousto-optical tunable filter (Crystal Technologies) with a bandwidth  of 1-2 nm~\cite{Sersic11b}. The beam is chopped for lock-in detection on an InGaAs photodiode. We polarize the incident beam using a broadband quarter-wave plate, to provide   circularly polarized excitation.
We weakly focus the beam onto the sample ($f$=100 mm). Light is  collected with a low NA collection lens ($f$=20 mm), and passed through a telescope  and pinhole to ensure spatial selection from within a 200x200 $\mu$m$^2$ e-beam write field, as monitored by an InGaAs camera. A motorized rotation stage allows transmission measurements versus incident angle relative to the sample normal.

Fig.~\ref{measurements} shows transmission versus wavelength for left and right handed circularly polarized incident light, for   incidence angles  from -50$^\circ$ to +50$^\circ$. Fig.~\ref{measurements} (a) shows data when  the   angle is varied from normal incidence by rotating the SRRs around their mirror axis $y$. At normal incidence, the magnetic LC resonance is evident around 1600 nm wavelength as a minimum in transmission. As opposed to the deep minima usually reported for linear, $x$-polarized transmission ($E$ along the gap) of dense arrays, the transmission dip is shallow since our lattice is dilute and the  LC resonance is associated only with $E_x$ and $H_z$, and completely transparent for $E_y$. 
As the incidence angle is moved away from the normal, the excitation also offers $H_z$ as a driving field, a quarter wave out of phase with $E_x$. A very clear asymmetry around the normal develops.  For right-handed light the transmission minimum becomes continuously shallower towards negative angles, and the sample is nearly transparent for $-50^\circ$. In contrast, the transmission minimum  significantly deepens from 28\% to 75\% when going to large positive angles. The asymmetric behavior with incidence angle is mirrored for opposite handedness (Fig.~\ref{measurements}(b), consistent with oblique incidence optical activity.  For linear polarization  the transmission is  symmetric around normal incidence (not shown).

The fact that optical activity is   symmetry-allowed even for lattices containing 2D non-chiral objects aligned with the lattice symmetry, was already reported by Plum et al.~\cite{Plum11}, who coined this `extrinsic 3D chirality'. In contrast to symmetry arguments that only distinguish between allowed and forbidden effects without quantifying the strength of optical activity, it is the express aim in this Letter to ascertain what the single element polarizability is that leads to the strong optical activity. We exclude the array structure factor as the cause of handed behavior~\cite{Gompf11}, as the optical activity disappears when  we rotate the SRRs by 90$^\circ$ in the sample plane (Fig.~\ref{measurements} (c) and (d)).   We hence conclude that the single SRR polarizability must contain the   strong `pseudo-chiralilty' that is expressed as  huge circular dichroism contrast in the extinction cross section, despite SRRs being  neither  2D nor 3D chiral.  Qualitatively, the  $LC$ description of  a single SRR  indeed contains optical activity under oblique incidence. Charge motion is set by  $q=(i\omega L +R+1/C)^{-1} [ i \omega\mu_0 A H_z + E_x/t]$, where $L$ is the inductance, $C$ the capacitance, $R$ the Ohmic resistance, $t$   the capacitor plate gap and $A$  the enclosed area. Full transparency despite the presence of suitable driving $E_x$  along the gap and $H_z$ through the split ring occurs when  $i \omega\mu_0 A H_z = -  E_x/t$. Conversely, optimum driving of a SRR benefits from an opposite quarter wave phase difference between $E_x$ and $H_z$ so that $ [ i \omega\mu_0 A H_z + E_x/t]$ has maximum magnitude.     Circular polarization at oblique incidence  provides the required quarter wave phase difference between $E_x$ and $H_z$.

\begin{figure}
\includegraphics[width=\columnwidth]{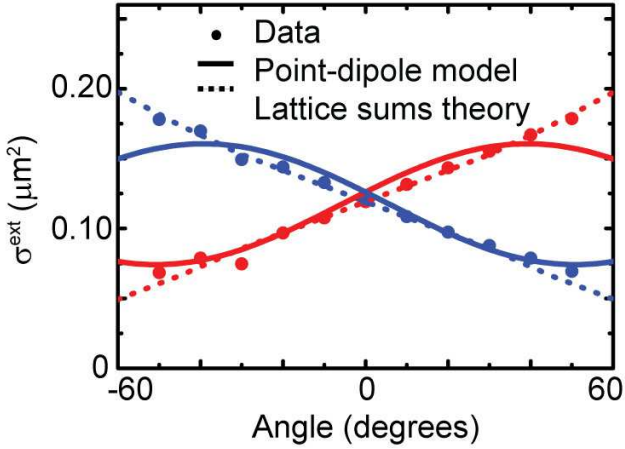}
\caption{Circles: effective extinction per  SRR from transmission data. Solid line:   single scatterer extinction cross section expected in a dipole model.  Dashed line: lattice sum calculation for a square array  with pitch  $d$=530 nm of magneto-electric dipoles.} \label{calculations}
\end{figure}

We quantify the polarizability tensor   from  the data by analyzing the effective extinction cross section per SRR defined as $\sigma=(1-T)d^2$, where $d$ is the lattice pitch and $T$ is the minimum in transmission~\cite{Sersic09}.  Fig.~\ref{calculations} shows that this  \emph{effective} extinction cross section  varies between 0.07 and  $0.16~\mu$m$^2$ as the angle is swept from $\pm50^\circ$ to  $\mp 50^\circ$  (mirrored dependence for opposite handedness).    For a single  magneto-electric   dipole scatterer Ref.~\cite{Sersic11a} predicts that the extinction  cross section  generally depends on angle $\theta$ as
\begin{equation}
\sigma_{R,L} (\theta)= \sigma_{-} + (\sigma_{+} -\sigma_{-})[1+\cos(2(\theta\pm\theta_0))]/2.\label{eq:angledep}
\end{equation}
 Measurements on a single object would provide the electrodynamic~\cite{Palik} $\alpha_E$ through  the normal incidence extinction $\sigma_{R,L}(0) = 2\pi k \mathrm{Im} \alpha_E$, while   the maximum and minimum attained  extinction  $\sigma_\pm$ encode electrodynamic polarizability eigenvalues via $\sigma_\pm=\pi k\mathrm{Im}(\alpha_E+\alpha_H\pm \sqrt{(\alpha_E-\alpha_H)^2+4\alpha_C ^2})$.  Such a fit of the single object extinction to  the measured effective extinction would provide $\alpha_E =4.1V$, $\alpha_H=3.6V$ and $\alpha_C=1.4V$ expressed in units of the geometrical volume of the SRR ($V=0.0012~\mu$m$^3$).  However, in a lattice of SRRs, the response is modified by coherences such that
$(\mathbf{p},\mathbf{m})^T=1/[ \bm{\alpha}^{-1}-{\cal{G}} ](\mathbf{E},\mathbf{H})^T$, where a lattice sum Green function
${\cal{G}}$ renormalizes the polarizability~\cite{Abajo07}. We calculate lattice transmission by rigorous electrodynamic lattice sums involving all multiple-scattering interactions between SRRs~\cite{Abajo07}. Consistent with our data, the calculated  transmission  shows strong optical activity under oblique incidence. We extract $\alpha_E=6.4V$, $\alpha_H=0.9V$ , $\alpha_C=2.1V$  at $\lambda$=1600 nm from a comparison to data, highlighting that the response of SRR arrays is consistent  with remarkably strong maximum magneto-electric cross coupling.

In Ref.~\cite{Sersic11a}  we analyzed how electrodynamic scatterers with arbitrary   polarizabilities of the form in Eq.~(\ref{eq:generalform}) scatter.  In that work, we realized that once one applies the optical theorem  to a planar scatterer (in-plane $\mathbf{p}$, out-of-plane $\mathbf{m}$),  $\bar{\alpha}_C \leq \sqrt{\bar{\alpha}_E \bar{\alpha}_H}$  appears as the maximum value that $\bar{\alpha}_C$  -- the crosscoupling after taking  a common resonant frequency factor  out of Eq.~(\ref{eq:generalform})~\cite{footnote} -- can possibly attain to avoid violation of energy conservation.
Here we claim that \emph{any} planar circuit-derived scatterer  is necessarily  exactly at this upper bound, i.e., at maximum cross coupling.  To prove this assertion we analyze a generic model for the polarizability  of a planar  scatterer under two general assumptions: (1) a linear response and (2) that an  electric and magnetic dipole response  originate  from the \emph{same}  equation of motion for charge $q$ moving through the scatterer.
Linear response implies $q=C_E(\omega) E + C_H(\omega) H$, where $E$ ($H$) is in the plane (perpendicular to the plane) of the scatterer.
Since $\mathbf{p}$ and $\mathbf{m}$ both derive from the same charge motion,  $\bm{p}=A_p  q$ and $\bm{m}=A_m  \dot{q}=i\omega A_m  (\omega) q$, where $A_p$ and $A_m$ are  geometry-dependent constants. One now finds the electrostatic circuit polarizability  as
\begin{equation}
\bm{\alpha}_0=\begin{pmatrix}  A_p C_E(\omega) & A_p C_H(\omega) \\ i \omega A_m C_E(\omega)  &   i \omega A_m C_H(\omega) \\ \end{pmatrix}.
\end{equation}
For reciprocal materials,  Onsager's  relations constrain $\mathbf{\alpha}_E$ and $\alpha_H$ to be symmetric, as well as requiring  $A_p C_H (\omega)= -i\omega A_m C_E(\omega)$.  Taking out  a common frequency factor ${\cal{L}}(\omega)\propto C_E(\omega)$ that describes the circuit resonance,
one finds that $\alpha_0$  always take the form~\cite{footnote}
\begin{equation}
\bm{\alpha}_0={\cal{L}}(\omega)  \begin{pmatrix}  \bar{\alpha}_E &   i\omega\sqrt{\bar{\alpha}_E\bar{\alpha}_H} \\ -  i\omega\sqrt{\bar{\alpha}_E\bar{\alpha}_H}& \omega^2\bar{\alpha}_H \\ \end{pmatrix}.
\label{alphagen}
\end{equation}  The surprise is that Onsager constraints
\emph{leave no freedom} to choose the off-diagonal coupling $\bar{\alpha}_C$. Any  planar circuit element is  cross coupled, with cross coupling  $\bar{\alpha}_C=\sqrt{\bar{\alpha}_E \bar{\alpha}_H}$. Combining this finding with our result from Ref.~\cite{Sersic11a} we conclude that \emph{any} planar circuit-derived scatterer is not just cross coupled, but that this coupling is at the maximum cross coupling limit. Maximum cross coupling means one vanishing eigenpolarizability $\alpha_-=0$,  hence  complete transparency of the scatterer for one handedness under oblique incidence, which means huge optical activity contrast.

\begin{figure}
\includegraphics[width=\columnwidth]{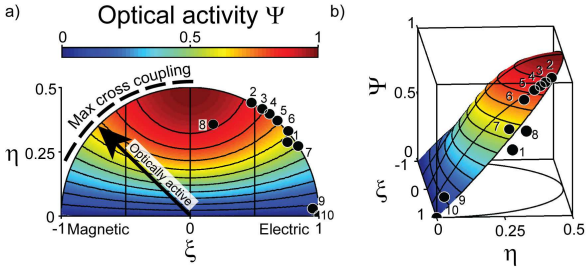}
\caption{Master diagrams summarizing optical activity and bi-anisotropy mapped as a function of $\xi=(\alpha_E-\alpha_H)/((\alpha_E+\alpha_H)$ and $\eta=(\alpha_C)/((\alpha_E+\alpha_H)$ . All structures we  tested   (data-points, numbered as in Fig.~\ref{cartoon}(b)) are close to the  locus of maximum cross coupling (ellipse), except (8). The color scale shows optical activity contrast $\Psi$,  in the dipole approximation (color scale) and for  tested structures (dots). Panel
(b) is a 3D representation of (a).}
\label{masterplot}
\end{figure}

Based on our experiment, we can now  assess whether the strong cross coupling in real scatterers is indeed close to the predicted maximum. From  the  polarizability we extracted from the  very strong circular polarization contrast in  extinction observed for split rings in~Figure~\ref{calculations}   we indeed find almost maximum cross coupling, since $\alpha_C \approx  0.88  \sqrt{\alpha_E \alpha_H}$. Furthermore, we  use full-wave simulations to examine the polarizability, and pseudo-chirality in extinction of many scatterers.  We use    3D Surface Integral Equation (SIE) calculations~\cite{Kern09}, to obtain full-wave solutions for archetypical metamaterial scatterers including SRRs, Omega particles with straight legs of different length,  double SRRs and double-gap rings as shown in Fig.~\ref{cartoon}(b). We calculate scattering cross sections and polarizability tensors independently from each other. To extract the polarizability, we excite  the same scatterer with six linearly independent illumination conditions, obtained as counter-propagating linearly polarized beams  set in (out of) phase to yield just electric (magnetic) Cartesian excitation.  We project the calculated scattered E-field evaluated on a spherical surface around the scatterer on vector spherical harmonics to retrieve $\mathbf{p}$ and $\mathbf{m}$~\cite{Muhlig11}.  As a consistency check on the polarizability retrieved by matrix inversion we verify that the Onsager  constraints are satisfied, which are not a priori assumptions in the retrieval.
We summarize results for all scatterers in a `master plot'  that allows comparison  independent of scatterer size. The scatterers are shown in Fig.~\ref{cartoon} (b). As a first dimensionless variable we use    $\xi=(\alpha_E - \alpha_H)/(\alpha_E+\alpha_H)$, which equals $\pm1$ for purely  electric  (magnetic) scatterers,     and $0$ for equal electric and magnetic polarizability.  As a dimensionless second variable we take  the normalized cross coupling $\eta=\alpha_C/(\alpha_E+\alpha_H)$. The locus of maximum cross coupling is the ellipse $\eta=\sqrt{1-\xi^2}/2$.  Fig.~\ref{masterplot} shows that most metamaterial scatterers we analyzed have $\xi$ well away from $1$, indicating significant magnetic polarizability. Furthermore  all   particles are essentially on the locus of maximum cross coupling, confirming our claim that bianisotropy is ubiquitous.

As third axis for the master plot we use a measure for  optical activity in scattering. All scatterers we simulated show an angular dependence of the scattering cross section of the  form in Eq.~(\ref{eq:angledep}). The dimensionless parameter $\Psi=|\sigma_{R}- \sigma_{L} |/(\sigma_{R}+ \sigma_{L} )$ evaluated at 45$^\circ$ incidence angle quantifies the maximum attained difference in extinction  $|\sigma_{R}- \sigma_{L}| $  (maximal always at 45$^\circ$) normalized to  (twice) the angle-averaged extinction cross section  $\sigma_++\sigma_-$.  Fig.~\ref{masterplot} shows $\Psi$  versus $\xi$ and $\eta$ as predicted by point scattering theory.  Evidently, optical activity is expected to be absent for zero cross coupling, and to increase  monotonically as cross coupling increases.   Very strong contrast in extinction per-building block is expected along most of the   locus of maximum cross coupling, vanishing only for purely electric, and purely magnetic dipole scatterers ($\xi=\pm 1$).  The full-wave simulations show that all the commonly used metamaterial scatterers exhibit strong optical activity in   surprisingly good agreement with the dipole model given that the circuit approximation, and the neglect of multipoles and retardation in Eq.~\ref{alphagen} are very coarse assumptions. Freedom to deviate significantly from the dipole model requires multiple overlapping resonances in a single scatterer. Indeed, the most noted deviations occur for the object (8,9) which has two hybridized resonances of   separate  parts. Earlier findings based on symmetry arguments  proposed that  extrinsic 3D chirality  requires loss~\cite{Plum11}. We find that optical activity  is in fact ubiquitous for planar magneto-electric scatterers, irrespective of absorption.  The cancellation  of optical activity  for zero absorption  noted by~\cite{Plum11} does not occur in $\bm{\alpha}$ but occurs in special cases where observables are subject to  additional symmetries, such as  wave vector conservation in  non-diffracting periodic systems.

To conclude, we have shown that  planar metamaterial scatterers that rely on a single resonance to generate a simultaneous electric and magnetic response are  maximally bianisotropic and strongly optically active, whether they  exhibit geometrical chirality or not.  Our findings have   important implications for controlling bianisotropy independently of $\epsilon$ and $\mu$ in metamaterials, since they imply that it is fundamentally impossible to independently control bianisotropy for single resonant objects. The only route to avoid bianisotropy in lattices of resonators is to   use  heterogeneous lattices that contain distinct, or multi-resonant elements (e.g., double-split rings in Fig.~\ref{masterplot}) to independently generate $\epsilon$ and $\mu$,  or to use lattices of effectively  larger `super-cells' with rotated copies of the same building block to cancel  off-diagonal coupling. 
Our results also hold important promise for  enhancing far-field or near-field chirality~\cite{Cohen10} in scattering applications where it is desired. In general, since maximum cross coupling is ubiquitous, optical activity  is  a very robust phenomenon that is easily extended to, e.g.,  finite clusters, random assemblies, or multi-element antennas. For instance, we predict that one can create chiral variants of the plasmon Yagi-Uda antenna to generate or selectively enhance circularly polarized single emitters.

\begin{acknowledgments}
We thank Huib Bakker and Ad Lagendijk for fruitful discussions. This work is part of the research program of the ``Stichting voor Fundamenteel
Onderzoek der Materie (FOM),'' which is financially supported by the
``Nederlandse Organisatie voor Wetenschappelijk Onderzoek (NWO).'' AFK acknowledges a VIDI fellowship funded by NWO.
\end{acknowledgments}


\begin{thebibliography}{99}
%negative refractive index
\bibitem{Veselago68}
V. G. Veselago, Sov.Phys. USPEKHI \textbf{10}, 509 (1968).

%%superlens
\bibitem{Pendry00}
J. B. Pendry, Phys. Rev. Lett. \textbf{85}, 3966 (2000);
J. B. Pendry, Physics World \textbf{14}, 47 (2001); C. M.
Soukoulis, S. Linden, and M. Wegener, Science \textbf{315}, 47
(2007); V. M. Shalaev, Nature Photonics \textbf{1}, 41 (2007).

%transformation optics
\bibitem{Pendry06}
J. B. Pendry, D. Schurig, and D. R. Smith, Science \textbf{312}, 1780 (2006).
\bibitem{Leonhardt06}
U. Leonhardt, Science \textbf{312}, 1777 (2006).
\bibitem{Schurig06}
D. Schurig, J. J. Mock, B. J. Justice, S. A. Cummer, J. B. Pendry, A. F. Starr, and D. R. Smith, Science \textbf{314}, 977 (2006).
%negative reflection

\bibitem{Zhou09}
J. Zhou, J. Dong, B. Wang, T. Koschny, M. Kafesaki, and C. M. Soukoulis, Phys. Rev. B. \textbf{79}, 121104 (2009).
\bibitem{Li10}
Z. Li, R. Zhao, T. Koschny, M. Kafesaki, K. B. Alici, E. Colak, H. Caglayan, E. \"{O}zbay, and C. M. Soukoulis, Appl. Phys. Lett. \textbf{97}, 08190 (2010).

%G's
\bibitem{Valev09}
V. K. Valev, N. Srnisdorn, A. V. Silhanek, B. De Clercq, W. Gillijns, M. Ameloot, V. V. Moshchalkov, and T. Verbiest, Nano Lett. \textbf{9}, 3945 (2009).
%nanoaperture
\bibitem{Gorodetsky09}
Y. Gorodetsky, N. Shitrit, I. Bretner, V. Kleiner, and E. Hasmanm Nano Lett. \textbf{9}, 3016 (2009).
\bibitem{Drezet08}
A. Drezet, C. Genet, J.-Y. Laluet, and T. W. Ebbesen, Opt. Express \textbf{16}, 12559 (2008).

\bibitem{Zhang09}
S. Zhang, Y.-S. Park, J. Li, X. Lu, W. Zhang, and X. Zhang, Phys. Rev. Lett. \textbf{102}, 023901 (2009).
\bibitem{Decker10}
M. Decker, R. Zhao, C. M. Soukoulis, S. Linden and M. Wegener, Opt. Lett. \textbf{35}, 1593 (2010).
\bibitem{Plum09}
E. Plum, J. Zhou, J. Dong, V. A. Fedotov, T. Koschny, C. M. Soukoulis, and N. I. Zheludev, Phys. Rev. B. \textbf{79}, 035407 (2009).
\bibitem{Gansel09}
J. K. Gansel, M. Thiel, M. S. Rill, M. Decker, K. Bade, V. Saile, G. von Freymann, S. Linden, and M. Wegener, Science \textbf{325}, 1513 (2009).
%\bibitem{Zhang07}
%C. Zhang and T. J. Cui, Appl. Phys. Lett. \textbf{91}, 194101 (2007).
\bibitem{Pendry04}
J. B. Pendry, Science \textbf{306} 1353 (2004).
\bibitem{SoukoulisCasimir} R. Zhao, J. Zhou, Th. Koschny, E. N. Economou and C. M. Soukoulis,  Phys. Rev. Lett. \textbf{103}, 103602 (2009).
%chirality
\bibitem{Govorov10}
Z. Fan and A. O. Govorov, Nano Lett. \textbf{10}, 2580 (2010).
\bibitem{Hendry10}
E. Hendry, T. Carpy, J. Johnston, M. Popland, R. V. Mikhaylovskiy, A. J. Lapthorn, S. M. Kelly, L. D. Barron, N. Gadegaard,
and M. Kadodwala, Nature Nanotech. \textbf{5}, 783 (2010).
\bibitem{Cohen10}
Y. Tang and A. E. Cohen, Phys. Rev. Lett. \textbf{104}, 163901 (2010).
\bibitem{Cohen11}
Y. Tang and A. E. Cohen, Science \textbf{332}, 333 (2011).

%\bibitem{Shivola07}
%A. Shivola, Metamaterials \textbf{1}, 2 (2007).

\bibitem{SihvolaBook} I. V. Lindell, A. H. Sihvola, S. A. Tretyakov, and A. J. Viitanen,
\emph{Electromagnetic Waves in Chiral and Bi Isotropic Media} (Artech
House, Norwood, MA, 1994).

\bibitem{SoukoulisAPL}N. Katsarakis, T. Koschny, M. Kafesaki, E. N. Economou, and C. M. Soukoulis, Appl. Phys. Lett. \textbf{84}, 2943 (2004).



\bibitem{Sersic11a}
I. Sersic, C. Tuambilangana, T. Kampfrath and A. F. Koenderink, Phys. Rev. B \textbf{83}, 245102 (2011). The appendix specifies the units also used in this Letter.


\bibitem{Kern09}
A. M. Kern and O. J. F. Martin, J. Opt. Soc. Am.  A \textbf{26}, 732 (2009). We use tabulated optical constants for gold~\cite{Palik}, and the following dimensions: inner/outer  radii in $\mu$m 0.74/1.19 (2-6), 1.6/2.5 and 2.7/3.6 (8), 2.7/3.6 (10), with a gap of 450 nm resp 200 nm for structures (2-6) resp. (10).  For scatterers (2-6) we increased the outer arm length from  0  to 900 nm. Scatterer thickness is 30 nm throughout. The respective resonance wavelength of the scatterers in $\mu$m are 1.600, 15.40, 16.06, 16.41, 16.80, 17.58, 1.544, 62.50,  23.25, and 16.50 for structures (1-10). Note that resonances (8,9) are two resonances in one structure.


\bibitem{footnote}In the static limit $\alpha_E, \alpha_H$ and $\alpha_C$ are real in ${\cal{A}}$ and Ohmic loss appears in ${\cal{L}}$ in Eq.~(\ref{alphagen}). The addition of radiation damping sets $\alpha_E$, $\alpha_H$ and $\alpha_C$ to be complex quantities even in absence of Ohmic damping~\cite{Sersic11a}.

\bibitem{Enkrich05} C. Enkrich, M. Wegener, S. Linden, S. Burger, L. Zschiedrich, F. Schmidt, J. F. Zhou, Th. Koschny, and C. M. Soukoulis, Phys. Rev. Lett. \textbf{95}, 203901 (2005).

\bibitem{Sersic09}
I. Sersic, M. Frimmer, E. Verhagen and A. F. Koenderink, Phys. Rev.
Lett. \textbf{103}, 213902 (2009).

\bibitem{Sersic11b}
I. Sersic, C. Tuambilangana and A. F. Koenderink, New J. Phys. \textbf{13}, 083019 (2011).

\bibitem{Plum11}
E. Plum, V. A. Fedotov, and N. I. Zheludev, J. Opt. \textbf{13}, 024006 (2011).


\bibitem{Gompf11}
B. Gompf, J. Braun, T. Weiss, H. Giessen, U. H\"ubner and M. Dressel, Phys. Rev. Lett. \textbf{106}, 185501 (2011).

\bibitem{Palik}We define a dynamic $\alpha$ by adding radiation damping $\mathrm{inv} (\alpha) = \mathrm{inv} (\alpha_{\mathrm{0}}) - i (2/3) k^3 \mathbb{I}$  to a static $\alpha_0$~\cite{Sersic11a} with Lorentzian  resonance ${\cal{L}}(\omega)$ centered at 1600 nm, and with  damping rate of gold $\gamma=1.25\cdot10^{14} s^{-1}$ from  \textit{Handbook of Optical Constants of Solids}, edited
by E. D. Palik (Adacemic, Orlando, FL, 1985).


\bibitem{Abajo07}
F. J. Garc\'{i}a de Abajo, Rev. Mod. Phys. \textbf{79}, 1267 (2007).



\bibitem{Muhlig11}
S. M\"uhlig, C. Menzel, C. Rockstuhl, and F. Lederer, Metamaterials \textbf{5}, 64 (2011).


\end{thebibliography}
\end{document}